\documentclass[aps,amsmath,notitlepage,twocolumn,amssymb,prl,longbibliography]{revtex4-1}

\usepackage{graphicx}
\usepackage{dcolumn}
\usepackage{bm}
\usepackage[colorlinks=true,citecolor=red,urlcolor=blue]{hyperref}
\usepackage{wasysym}
\usepackage{stmaryrd}
\usepackage{verbatim}
\usepackage{subfigure}

\begin{document}
\title{Unusual double-peak specific heat and spin freezing in a spin-2 \\triangular lattice antiferromagnet FeAl$_2$Se$_4$}

\author{Kunkun Li$^{1,2}$}
\author{Shifeng Jin$^{1,3}$}
\author{Jiangang Guo$^{1}$}
\author{Yanping Xu$^{1}$}
\author{Yixi Su$^{5}$}
\author{Erxi Feng$^{5}$}
\author{Yu Liu$^{6}$}
\author{Shengqiang Zhou$^{6}$}
\author{Tianping Ying$^{7}$}
\author{Shiyan Li$^{7,9,10}$}
\author{Ziqiang Wang$^{11}$}
\email{wangzi@bc.edu}
\author{Gang Chen$^{7,8,10}$}
\email{gangchen.physics@gmail.com}
\author{Xiaolong Chen$^{1,3,4}$}
\email{chenx29@iphy.ac.cn}

\affiliation{$^1$Research \& Development Center for Functional Crystals,
Beijing National Laboratory for Condensed Matter Physics, Institute of Physics,
Chinese Academy of Sciences, Beijing 100190, China}
\affiliation{$^2$University of Chinese Academy of Sciences, Beijing, 100049, China}
\affiliation{$^3$School of Physical Sciences, University of Chinese Academy of Sciences, Beijing 101408, China}
\affiliation{$^4$Collaborative Innovation Center of Quantum Matter, Beijing, China}
\affiliation{$^5$Juelich Centre for Neutron Science (JCNS) at Heinz Maier-Leibnitz Zentrum (MLZ),
Forschungszentrum Juelich GmbH, Lichtenbergstr. 1, 85747 Garching, Germany}
\affiliation{$^6$Helmholtz-Zentrum Dresden-Rossendorf,
Institute of Ion Beam Physics and Materials Research,
Bautzner Landstra$\beta$e 400, 01328 Dresden, Germany}
\affiliation{$^7$State Key Laboratory of Surface Physics,
Department of Physics, Fudan University, Shanghai 200433, China}
\affiliation{$^8$Center for Field Theory and Particle Physics, Fudan University, Shanghai 200433, China}
\affiliation{$^9$Laboratory of Advanced Materials, Fudan University, Shanghai 200433, China}
\affiliation{$^{10}$Collaborative Innovation Center of Advanced Microstructures, Nanjing University,
Nanjing 210093, China}
\affiliation{$^{11}$Department of Physics, Boston College, Chestnut Hill, Massachusetts 02467, USA}

\date{\today}

\begin{abstract}
We report the properties of a triangular lattice iron-chalcogenide
antiferromagnet FeAl$_2$Se$_4$. The spin susceptibility reveals
a significant antiferromagnetic interaction with a Curie-Weiss
temperature ${\Theta_{\text{CW}} \simeq -200 \text{K}}$ and a
spin-2 local moment. Despite a large spin and a large
$|\Theta_{\text{CW}}|$, the low-temperature behaviors are
incompatible with conventional classical magnets. No long-range
order is detected down to 0.4K. Similar to the well-known
spin-1 magnet NiGa$_2$S$_4$, the specific heat of FeAl$_2$Se$_4$
exhibits an unusual double-peak structure and a $T^2$ power law at
low temperatures, which are attributed to the underlying quadrupolar
spin correlations and the Halperin-Saslow modes, respectively.
The spin freezing occurs at ${\sim 14}$K, below which the relaxation
dynamics is probed by the ac susceptibility. Our results are
consistent with the early theory for the spin-1 system with
Heisenberg and biquadratic spin interactions. We argue that the
early proposal of the quadrupolar correlation and gauge glass
dynamics may be well extended to FeAl$_2$Se$_4$. Our results
provide useful insights about the magnetic properties of
frustrated quantum magnets with high spins.
\end{abstract}

\maketitle

\emph{Introduction.}---Magnetic frustration arises in systems
with competing spin interactions that cannot be optimized simutaneously~\cite{Balents}
In general, sufficiently strong frustration could lead to degenerate
or nearly degenerate classical spin states and thus induce exotic and
unconventional quantum states of matter such as quantum spin liquids
when the quantum mechanical nature of the spins is considered. The
conventional wisdom and belief tells us that it is more likely to
find these quantum states in magnetic systems with spin-1/2 degrees
of freedom on frustrated lattices where quantum fluctuations are
deemed to be strong. This explains the major efforts and interests
in the spin-1/2 triangular lattice magnets like Cs$_2$CuCl$_4$~\cite{cscucl,cscucl2},
$\kappa$-(BEDT-TTF)$_2$Cu$_2$(CN)$_3$~\cite{kappaET,organics1,organictherm}, 
EtMe$_3$Sb[Pd(mit)$_2$]$_2$~\cite{dmit} and YbMgGaO$_4$~\cite{Yuesheng1,Yuesheng2,YaoShenNature,Martin2016,PhysRevB.96.054445,SciPostPhys.4.1.003,PhysRevLett.119.157201,PhysRevB.97.125105} 
the spin-1/2 kagom\'{e} lattice magnets like
herbertsmithite ZnCu$_3$(OH)$_6$Cl$_2$,
volborthite~\cite{voborthite} Cu$_3$V$_2$O$_7$(OH)$_2 \cdot 2$H$_2$O
and kapellasite~\cite{PhysRevLett.109.037208} 
Cu$_3$Zn(OH)$_6$Cl$_2$, various spin-1/2 rare-earth pyrochlore magnets~\cite{RevModPhys.82.53},
and other geometrically frustrated lattices with spin-1/2 moments or effective spin-1/2
moments~\cite{hyperk}. Despite the tremendous efforts in the spin-1/2 magnets,
the magnets with higher spin moments can occassionally be interesting.
The exceptional examples of this kind are the well-known Haldane phase~\cite{PhysRevLett.50.1153,PhysRevLett.59.799}
for the spin-1 chain and its high dimensional extension such as topological
paramagnets~\cite{Chen1604,PhysRevB.91.195131}. 
The former has been discovered in various Ni-based 1D 
magnets~\cite{PhysRevLett.56.371,PhysRevLett.63.1424,PhysRevLett.72.3108}.
Another well-known example is the spin-1 triangular lattice antiferromagnet~\cite{Nakatsuji1697,PhysRevLett.99.157203,PhysRevLett.105.037402} 
NiGa$_2$S$_4$, where the biquadratic spin 
interaction~\cite{PhysRevB.79.214436,PhysRevB.74.092406,doi:10.1143/JPSJ.75.083701}, 
that is completely
absent for spin-1/2 magnets, bring the spin quadrupolar order/correlation
(or spin nematic) physics and phenomena into the system. Therefore, what
matters is not just the size of the spin moment, but rather the interactions
among the local moments and the underlying lattices.

Inspired by the potentially rich physics in high-spin systems, in this Letter,
we study a spin-2 triangular lattice antiferromagnet FeAl$_2$Se$_4$ with both
polycrystalline and single crystalline samples.
Analogous to the Ni$^{2+}$
local moments in NiGa$_2$S$_4$~\cite{Nakatsuji1697,PhysRevLett.99.157203,PhysRevLett.105.037402},
 the Fe$^{2+}$ local moments in this material
form a perfect triangular lattice and provide a perfect setting to explore
the quantum physics of high spin moments on frustrated lattice. Unlike the
usual classical behaviors expected for high spins, we find that the Fe local
moments remain disordered down to 0.4K despite a rather large antiferromagnetic
Curie-Weiss temperature ${\Theta_{\text{CW}} \simeq -200}$K. The magnetic
susceptibility of single crystal samples show a bifurcation at about 14K
for field cooling and zero-field cooling measurements,
suggesting a glassy like spin freezing. This is further assured
from the ac susceptibility measurements at different probing frequencies.
The specific heat of FeAl$_2$Se$_4$ shows an unusual double-peak structure
at two well-separated temperatures, indicating two distinct physical processes
are occurring. Below the spin freezing temperature, a $T^2$ power law
specific heat is observed. Based on the early theoretical 
works~\cite{PhysRevB.79.214436,PhysRevB.79.140402,PhysRevB.74.092406,doi:10.1143/JPSJ.75.083701} on
NiGa$_2$S$_4$, we propose that the double-peak structure in heat capacity
arises from the growth of correlation of two distinct types of spin
moments, and the $T^2$ power law is the consequence of the Goldstone-type
spin waves (i.e. the Halperin-Saslow modes). We further suggest that the
spin freezing is due to the disorder that may induce the gauge glass physics
into the would-be ordered state of this system.

\begin{figure}[t]
\includegraphics[width=7cm]{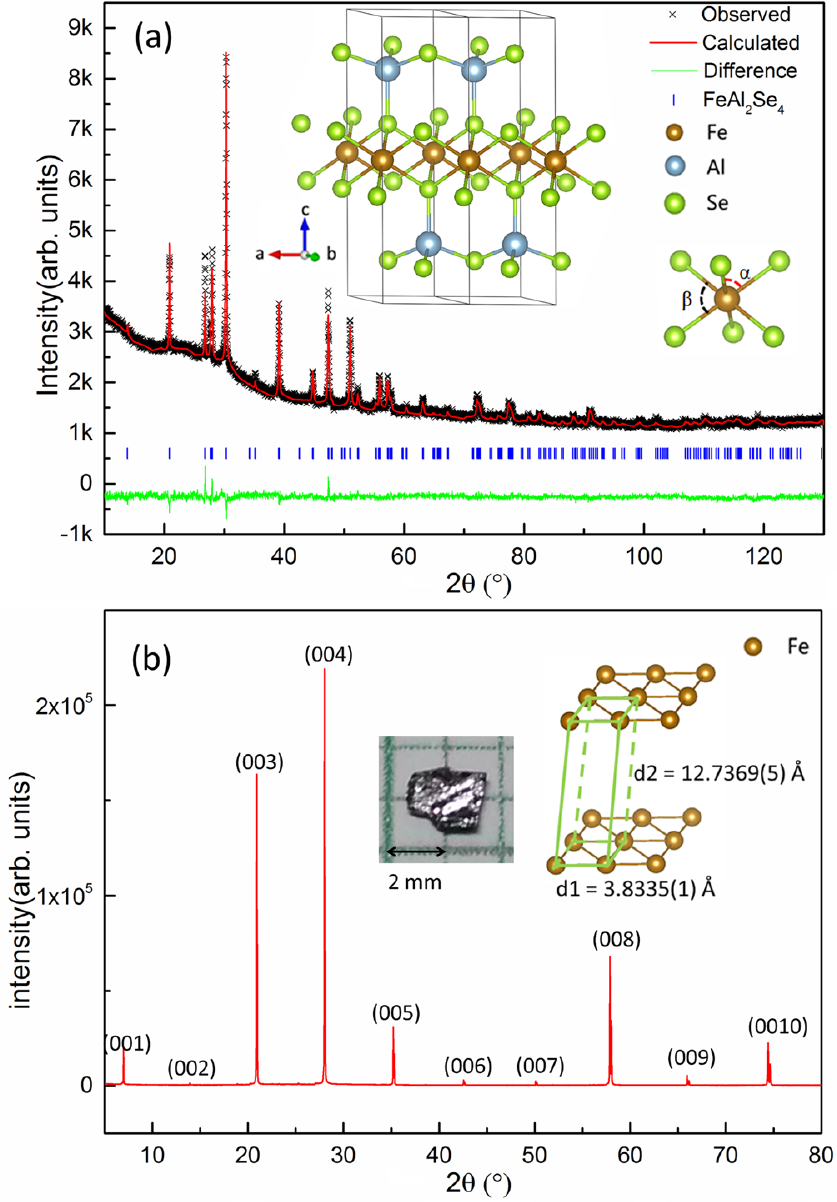}
\caption{(Color online.)
(a) Powder X-ray diffraction and Ritveld refinement profile of FeAl$_2$Se$_4$
at the room temperature. The inset shows the schematic crystal structure of FeAl$_2$Se$_4$
and the distorted FeSe$_6$ octahedra.
(b) The X-ray diffraction pattern of FeAl$_2$Se$_4$ crystal indicates
that the (00l) reﬂections dominate the pattern.
Inset shows the photography of the FeAl$_2$Se$_4$  crystal with a
length scale of 2 mm and an exhibition of the Fe sublattice.
}
\label{fig1}
\end{figure}

\emph{Crystal structure.}---Our polycrystalline and single crystal FeAl$_2$Se$_4$
samples were prepared from the high temperature reactions of high purity
elements Fe, Al and Se. In Fig.~\ref{fig1}(a), we show the room temperature
X-ray diffraction pattern on the powder samples that are obtained
by grinding the single crystal samples. All the reflections could
be indexed with the lattice parameters ${a = b = 3.8335(1)}$\AA,
${c = 12.7369(5)}$\AA, and the unit cell volume ${V = 162.108(0.01)}$\AA$^3$.
The systematic absences are consistent with space group P$\bar{3}$m1 (No.~164),
suggesting FeAl$_2$Se$_4$ is isostructural to the previously
reported compound NiGa$_2$S$_4$~\cite{Nakatsuji1697}. The structural parameters are listed in
Table S1 in the Supplementary material. In Fig.~\ref{fig1}(b), we show the X-ray
diffraction pattern of the single crystal samples. It clearly indicates
that the cleaved surface of the flaky crystal is the (001) plane and
normal to the crystallographic $c$ axis. The composition was further
examined by inductively coupled plasma atomic emission spectrometer,
giving the atomic ratios of {Fe: Al: Se} close to ${1:2:4}$.
The compound is built by stacking of layers consisting of
edge-sharing FeSe$_6$ octahedra that are connected by a top
and a bottom sheet of AlSe$_4$ tetrahedra. The layers are
separated with each other by a van der Waals gap. The central
FeSe$_6$ octahedra layer is isostructural to the CoO$_2$ layer
of the well-known superconducting material~\cite{NaxCoO2} 
Na$_x$CoO$_2 \cdot y$H$_2$O.

In the crystal field environment of FeAl$_2$Se$_4$, the Fe$^{2+}$ ion
has an electronic configuration $t_{2g}^4 e_g^2$ that gives
rise to a high spin state and a spin ${S = 2}$ local moment.
The six Fe-Se bonds are of equal length and are 2.609(3)\AA.
Se-Fe-Se angles are 94.55(8)$^{\circ}$, marked as $\alpha$,
and 85.45(8)$^{\circ}$, marked as $\beta$, as displayed in
the inset of Fig.~\ref{fig1}(b). The different Se-Fe-Se angles
represent a slight rhombohedral distortion of the FeSe$_6$ octahedra,
resulting in a small crystal field splitting among the $t_{2g}$ orbitals.
The degenerate or nearly degenerate $t_{2g}$ orbitals and the partially
filled $t_{2g}$ shell may lead to an active orbital degree of freedom.
This will be further discussed from the magnetic entropy measurement.

Finally, in FeAl$_2$Se$_4$, the nearest intralayer Fe-Fe
distance is ${d_1 = 3.8335(1)}$\AA\, and the nearest interlayer
Fe-Fe distance is ${d_2 = 12.7369(5)}$\AA, indicating an ideal
two-dimensional character in terms of the lattice structure.

\begin{figure}[t]
\includegraphics[width=7cm]{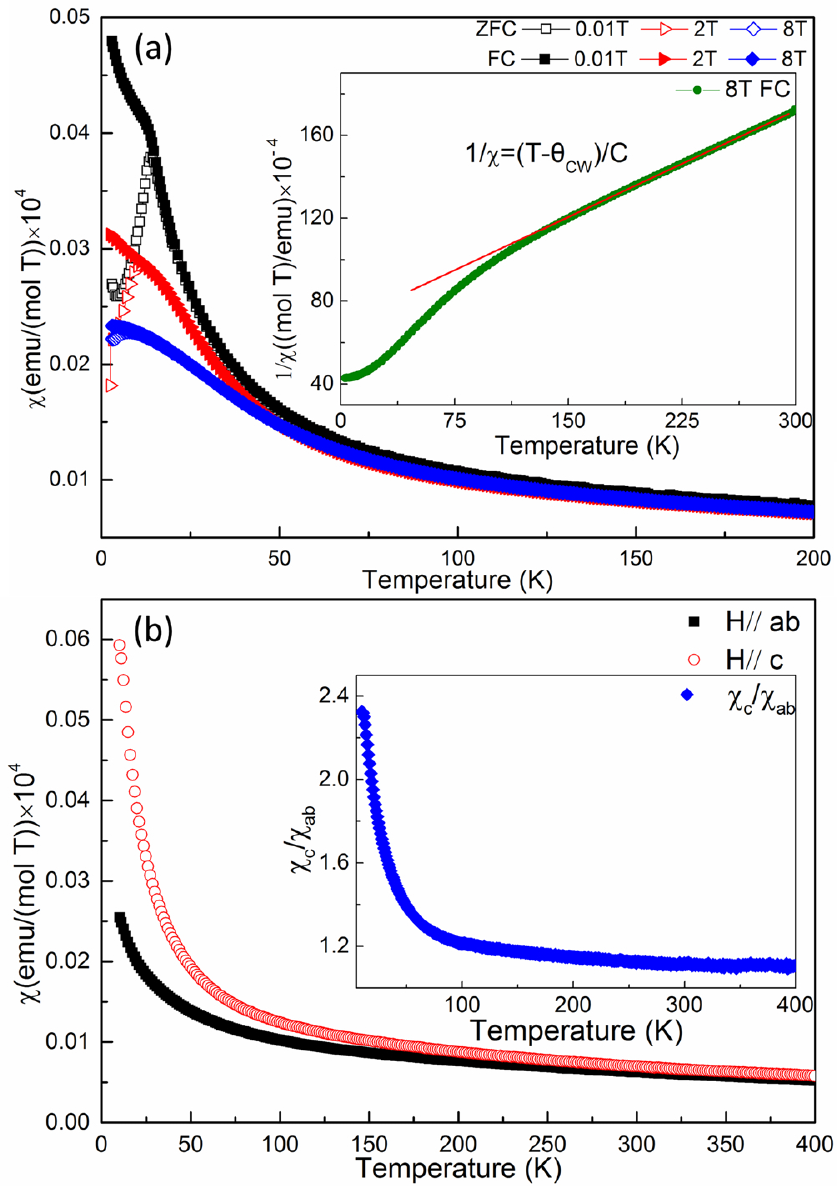}
\caption{(Color online.)
(a) Zero-field-cold (ZFC) and field-cold (FC) $\chi(T)$ data taken at
different applied ﬁeld from 2K to 300K. Inset: the inverse susceptibility
$\chi^{-1}(T)$ data with the applied field of 8T. The solid red lines
are linear fits with a Curie-Weiss law ${\chi = C/(T - \Theta_{\text{CW}})}$.
(b) Temperature dependence of the magnetic susceptibility $\chi_{ab}$
and $\chi_c$ obtained for FeAl$_2$Se$_4$ under 0.1T from 10K to 400K.
Inset: the ratio of $\chi_c/\chi_{ab}$ vs $T$.}
\label{fig2}
\end{figure}

\emph{Thermodynamic measurements.}---To identify the magnetic properties
of FeAl$_2$Se$_4$, we first implement the thermodynamic measurements.
The temperature dependent $dc$ magnetic susceptibility and its inverse
$\chi^{-1}$ under the external magnetic fields of 0.01T, 2T and 8T
are shown in Fig.~\ref{fig2}. A bifurcation (denoted as $T_f$) at 14K
can be seen under a field of 0.01T, and can be suppressed down to 8K
when the applied field is raised up to 8T. This is a signature of
spin freezing. The temperature dependent susceptibility from 150K
to 300K obeys a simple Curie-Weiss law ${\chi = C/(T - \Theta_{\text{CW}})}$,
where $C$ is the Curie constant and $\Theta_{\text{CW}}$ is the Weiss
temperature as illustrated in the inset of Fig.~\ref{fig2}(a).
The effective magnetic moments, 4.80-5.20$\mu_{B}$, were obtained from
the Curie constants. The Weiss temperature ${\Theta_{\text{CW}} = -200}$K,
that is more negative than that for the isostructural material~\cite{PhysRevLett.99.157203}
FeGa$_2$S$_4$ (${\Theta_{\text{CW}} = -160}$K), indicates stronger
antiferromagnetic interactions. When the temperature is lower than
150K, FeAl$_2$Se$_4$ shows a deviation from the Curie-Weiss behavior.
The frustration index, defined by ${f = |\Theta_{\text{CW}}/T_f|}$
with $T_f$ the spin freezing temperature, is estimated as 14.
This is a relatively large value, and we thus conclude that
FeAl$_2$Se$_4$ is a magnetically frustrated system. In Fig.~\ref{fig2}(b)
we further show the magnetic susceptibility measurements of the
FeAl$_2$Se$_4$ single crystals with ${H \parallel ab}$ ($\chi_{ab}$)
and ${H \parallel c}$ ($\chi_c$). Unlike NiGa$_2$S$_4$ and FeGa$_2$S$_4$~\cite{PhysRevLett.99.157203},
here we find an easy-axis anisotropy with $\chi_c/\chi_{ab}$ about 2.4
at 5K instead of easy-plane anisotropy. This is due to the partially
filled $t_{2g}$ shell in the Fe$^{2+}$ ion where the spin-orbit coupling
is active and induces the anisotropy in the spin space.

\begin{figure}[t]
\includegraphics[width=7cm]{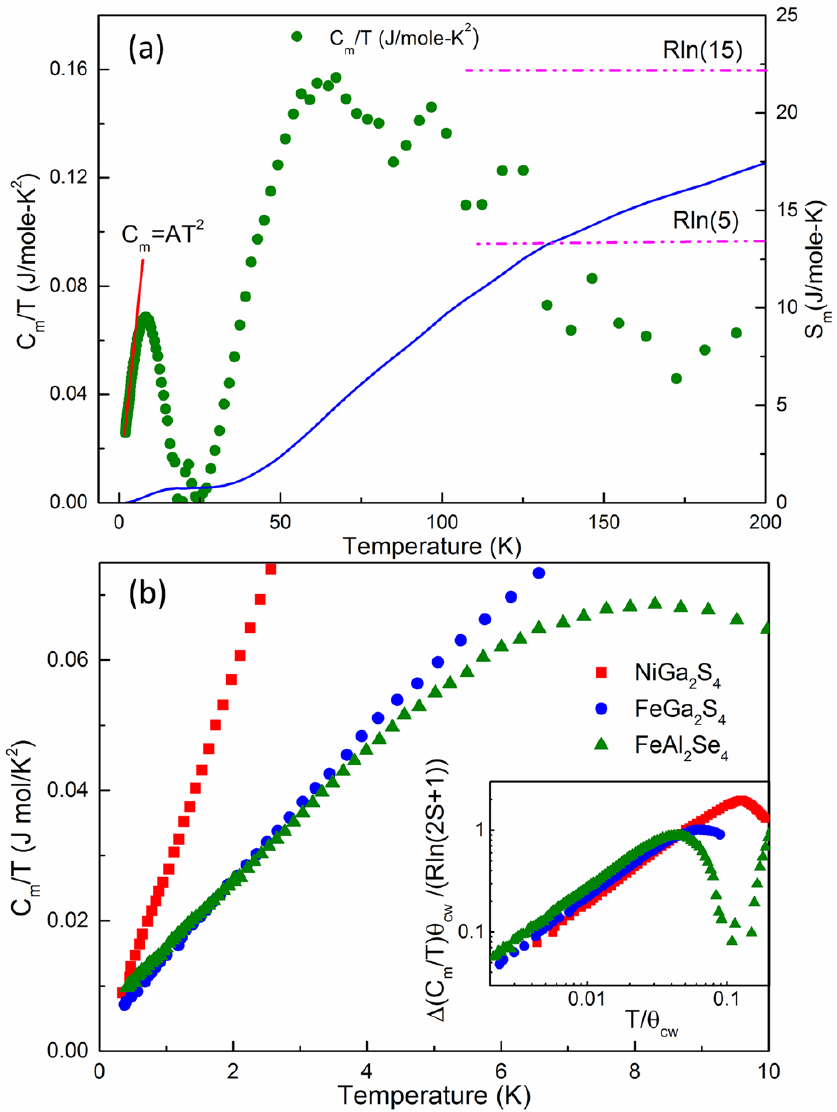}
\caption{(Color online.)
(a) Temperature dependence of magnetic entropy (right axis) and $C_m/T$
(left axis) for FeAl$_2$Se$_4$. (b) The low-temperature part of magnetic
heat capacity $C_m/T$ for NiGa$_2$S$_4$, FeGa$_2$S$_4$ and FeAl$_2$Se$_4$.
Inset shows the $\Delta(C_m/T)\Theta_{\text{CW}}/[R\ln (2S+1)] $ vs
$T/\Theta_{\text{CW}}$ for NiGa$_2$S$_4$ (${S = 1}$,
${\Theta_{\text{CW}} =- 80}$K),
FeGa$_2$S$_4$ (${S = 2}$, ${\Theta_{\text{CW}} = -160}$K)
and FeAl$_2$Se$_4$ (${S = 2}$, ${\Theta_{\text{CW}} =- 200}$K)
at 0T in full logarithmic scale.}
\label{fig3}
\end{figure}

The magnetic heat capacity after subtracting the phonon contributions
is used to reveal the spin contribution. The heat capacity of
an isostructural non-magnetic material ZnIn$_2$S$_4$ is measured
to account for the lattice contribution of FeAl$_2$Se$_4$. No
clear anomaly associated with any magnetic transition can be
detected from the specific heat data down to 0.4K, indicating
a ground state without any true long range spin ordering.
The magnetic contribution $C_m$ is obtained by subtracting
the phonon contribution from ZnIn$_2$S$_4$. Similar to
NiGa$_2$S$_4$ and FeGa$_2$S$_4$~\cite{PhysRevLett.99.157203}, 
FeAl$_2$Se$_4$ exhibits a double-peak variation of $C_m/T$: 
one at ${\sim10}$K, and the other at ${\sim 65}$K,
as shown in Fig.~\ref{fig3}(a). We will revisit
the double-peak structure of the heat capacity later.

The magnetic entropy, ${S_m(T) = \int_0^T C_m/T \, dT}$,
increases gradually over the entire measured temperature
range but with a plateau near $T \sim 25$K, indicating high
degeneracy of low-energy states due to magnetic frustration.
The total entropy reaches $R\ln(5)$ at ${T \sim 135}$K,
corresponding to the value for the ${S = 2}$ system.
Then it further increases towards ${R\ln(15) = R\ln(5)
+ R\ln(3)}$. The latter term is from the orbital degree
of freedom due to two holes present in the $t_{2g}$ orbitals~\cite{PhysRevLett.99.157203}.
The low-temperature part of $C_m/T$, as shown in Fig.~\ref{fig3}(b),
displays a near linear $T$ dependence around 4K and then
deviates from the line with further increasing temperature,
similar to the behavior that was observed in NiGa$_2$S$_4$
and FeGa$_2$S$_4$. Besides, the linear-$T$ coefficient
$\gamma$ for $C_m/T$ of 5.9mJ/molK$^2$ at ${T \rightarrow 0}$K
can be obtained for FeAl$_2$Se$_4$, slightly larger than that
in FeGa$_2$S$_4$ (3.1mJ/molK$^2$). The observed $T^2$
specific heat can be attributed to the Halperin-Saslow modes~\cite{PhysRevB.16.2154}
in two dimensions that give a specific heat of the form
\begin{eqnarray}
C_m = N_{\text A} \frac{3\pi k_BV}{c} \big[ \zeta(3) \sum_j \big( \frac{k_B T}{\pi \hbar v_j} \big)^2 -
\frac{1}{L_0^2} \big],
\end{eqnarray}
where $V = \sqrt{3}a^2c/2$ is the unit-cell volume with $a$ the Fe-Fe
spacing, $L_0$ is the coherence length for the spin excitations
and $v_j$ is the velocity in the $j$-th direction. Using the experimental data on the
susceptibility $\chi(T \rightarrow 0) = 0.0025$emu/mol and
$C_m/T^2 = 0.010$J/molK$^3$, the estimated spin stiffness
$\rho_s = \chi(v/\kappa)^2 = 49.5 $K, where $ \kappa=g\mu_B/\hbar$,
which is larger than those obtained in FeGa$_2$S$_4$
($\rho_s = 35.8 $K) and NiGa$_2$S$_4$ ($\rho_s = 6.5 $K).
To further compare FeAl$_2$Se$_4$ with the other two counterparts,
in the inset of Fig.~\ref{fig3}(b) we show
$\Delta(C_m/T)\Theta_{\text{CW}}/[R\ln(2S+1)]$ vs $T/\Theta_{\text{CW}}$
for NiGa$_2$S$_4$ (${S = 1}$, ${\Theta_{\text{CW}} = -80}$K),
FeGa$_2$S$_4$ (${S = 2}$, ${\Theta_{\text{CW}} = -160}$K)
and FeAl$_2$Se$_4$ (${S = 2}$, ${\Theta_{\text{CW}} =-200}$K)
at zero field (0T) in full logarithmic scale, where $\Delta(C_m/T)\Theta_{\text{CW}}=C_m/T-\gamma$.
As we show in Fig.~\ref{fig3}(b), the low-temperature
data for FeAl$_2$Se$_4$ nearly collapse on top of
NiGa$_2$S$_4$ and FeGa$_2$S$_4$, indicating similar
Halperin-Saslow modes present in all three compounds.

\emph{Double-peak heat capacity.}---Here we discuss the origin of the double-peak
structure in the heat capacity of FeAl$_2$Se$_4$. As it was noted, such a double-peak
structure was first observed in the spin-1 magnet NiGa$_2$S$_4$~\cite{Nakatsuji1697}.
The theoretical studies have invoked a spin model with both Heisenberg and biquadratic exchange
interactions~\cite{PhysRevB.79.214436,PhysRevB.74.092406,doi:10.1143/JPSJ.75.083701}, 
where the biquadratic exchange interaction,
$-({\boldsymbol S}_i \cdot {\boldsymbol S}_j)^2$, arises from the spin-lattice
coupling. Since FeAl$_2$Se$_4$ is isostructural to NiGa$_2$S$_4$, we expect similar
model and interactions to apply. The presence of the biquadratic exchange allows the
system to access the spin quadrupole moments effectively and hence enhance the
quadrupolar correlation. In addition to the usual magnetic (dipole) moment $S^{\mu}$,
Both spin-1 and spin-2 moments support the quadrupole moments
${Q_{\mu\nu} = \frac{1}{2} (S^{\mu}S^{\nu}+S^{\nu}S^{\mu})
- \frac{1}{3} S(S+1)\delta_{\mu\nu} }$ with ${\mu=x,y,z}$.
Since the quadrupole and dipole moments are quite distinct and have different
symmetry properties, they ought to behave differently. Moreover, it is the
biquadratic interaction that directly couples the quadrupole moments of
different sites. It was then argued and shown numerically~\cite{PhysRevB.79.214436} 
that the system develops significant quadrupolar correlations at a distinct higher
temperature than the one associated with the rapid growth of the magnetic
correlations when the system is close to the quantum phase transition from
spiral (dipolar) spin order to quadrupolar order. These two temperature scales
associated with the rapid growth of magnetic and quadrupolar correlations result
in an unusual double-peak structure of the heat capacity. This physics is certainly
not unique to the spin-1 NiGa$_2$S$_4$. Based on the fact that FeAl$_2$Se$_4$
has an identical lattice structure and an even larger spin Hilbert space,
we expect the same mechanism to account for the double-peak heat capacity
in FeAl$_2$Se$_4$.

\begin{figure}[t]
\includegraphics[width=7cm]{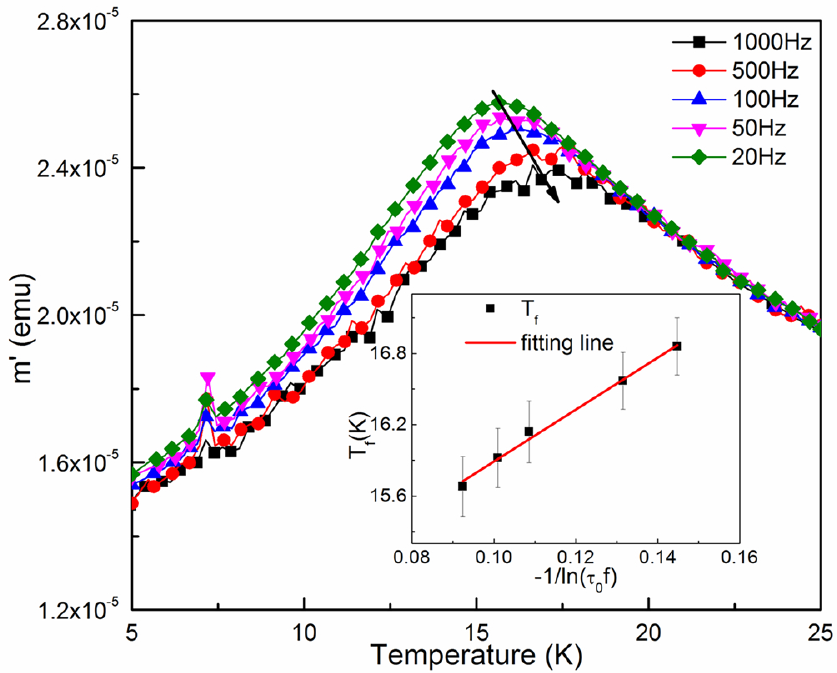}
\caption{(Color online.)
Temperature dependence of the real part of the ac magnetic susceptibility as a
function of frequency. It should be noted that the small peak emerging at
approximately 7K in the figure is frequency independent, the origin is unknown.
}
\label{fig4}
\end{figure}

\emph{Spin freezing and ac susceptibility.}---To further characterize the low-temperature
magnetic properties of FeAl$_2$Se$_4$ at temperatures near the spin freezing, we measure
the temperature dependent ac susceptibility from 5K to 25K for a number of frequencies.
As shown in Fig.~\ref{fig4}, a peak in the real part at $\sim$15 K is present, which is
the signature of the susceptibility bifurcation. A small but clear peak shift towards
high temperatures can be observed when the probing frequency is increased.
This suggests a spin relaxation behavior. The shift of the peak temperature
as a function of frequency described by the expression,
$(\Delta T_f)/(T_f \Delta \log \omega)$, is usually used to distinguish spin glass and
spin glass like materials~\cite{Glasses,PhysRevB.89.214401}. 
The value obtained for FeAl$_2$Se$_4$ is 0.042,
which is slightly larger than expected for a canonical spin glass but is in
the range of spin glass like materials. The Volger-Fulcher law is then applied
to characterize the relaxation feature with a function relating the bifurcation
temperature ($T_f$) with the frequency ($f$): $T_f=T_0 -  E_a  /[ k_B\ln(\tau_0 f) ]$,
where $\tau_0$ is the intrinsic relaxation time, $E_a$ the activation energy of
the process, and $T_0$ ``the ideal glass temperature''~\cite{0953-8984-21-10-105801}. 
The fitted $\tau_0$ is $1\times 10^{-7} $s, which is the same order as the 
super-paramagnets and cluster glasses. The activation energy of the process 
$E_a$ is 3.06meV.

Usually the spin freezing with the glassy behavior is due to disorder and/or
frustration that are present in FeAl$_2$Se$_4$. Like the S vacancies in
NiGa$_2$S$_4$, we suspect the Se vacancies to be the dominant type of
impurities and sources of disorder. Without disorders, the system may
simply develop the spin density or spiral magnetic orders. With
(non-magnetic bond) disorders, the phase transition associated with
the discrete lattice symmetry breaking would be smeared out in FeAl$_2$Se$_4$.
No sharp transition was observed in the heat capacity measurement
on FeAl$_2$Se$_4$. By assuming a complex XY order parameter for each magnetic 
domain in the spin freezing regime, the authors in Ref.~\onlinecite{PhysRevB.79.214436}
invoked a phenomenological gauge glass model~\cite{PhysRevB.38.386,PhysRevB.43.130} 
where the complex
orders from different magnetic domains couple with the disorder
in a fashion similar to the coupling with a random gauge link variable.
They propose that the system would realize a gauge glass ground state, 
and the Goldstone-type spin waves in a long-range ordered state turn 
into the Halperin-Saslow modes~\cite{PhysRevB.79.140402,PhysRevB.16.2154} 
in the gauge glass model. These gapless 
modes in two diemnsions contribute to the $T^2$ specific heat~\cite{PhysRevB.16.2154} 
in the spin freezing regime. Due to the phenomenological nature of the model, 
we think the gauge glass model and the conclusion should also describe and apply 
to the low-temperature physics in the spin freezing regime of FeAl$_2$Se$_4$.

\emph{Discussion.}---Although the large spin moments tend to behave more classically
than spin-1/2 moments, the large spin moments have a larger spin Hilbert space and
would allow more possibilities for the quantum ground states. If the interaction can
access these Hilbert space effectively, interesting quantum states may be stabilized.
From our experimental results in FeAl$_2$Se$_4$, we find that the system exhibits
two-peak structure in the heat capacity. We argue that these two peaks correspond to
the separate growth of quadrupolar correlation and magnetic (dipolar) correlation.
From the early experience with the spin-1 triangular lattice magnet NiGa$_2$S$_4$~\cite{Nakatsuji1697,PhysRevLett.99.157203,PhysRevLett.105.037402,PhysRevB.79.214436,PhysRevB.79.140402,PhysRevB.74.092406,doi:10.1143/JPSJ.75.083701},
we expect this physics is certainly not specific to FeAl$_2$Se$_4$, and it is not
even specific to spin-2 magnets nor to triangular lattice magnets. This type of physics, i.e.
the rich moment structure and their correlations, may broadly exist in frustrated magnets
with high spin moments if the interaction can access these high-order multipole moments effectively.
In the cases of FeAl$_2$Se$_4$ and NiGa$_2$S$_4$, it is the spin-lattice-coupling
induced biquadratic interaction that enhances the quadrupolar order and correlation.
Besides these two known examples, for other high spin systems such as the $4d$/$5d$ magnets~\cite{PhysRevB.82.174440,PhysRevB.84.094420}
and $4f$ rare-earth magnets~\cite{PhysRevLett.105.047201,PhysRevLett.112.167203,PhysRevB.94.201114,nematics}, 
the spin-orbit coupling and entanglement could
induce strong multipolar interaction and provide another
mechanism to access and enhance the quadrupolar (and more generally multipolar) spin orders
and correlations. Thus, we think our results and arguments could stimulate interests
in frustrated magnets with high spins and rich moment structures.

To be specific to FeAl$_2$Se$_4$, there are a couple directions for future experiments.
Since the biquadratic interaction is suggested to arise from the spin-lattice coupling,
it is useful to substitute some Se with S to modify the spin-lattice coupling and
hence the biquadratic interaction. This should affect the quadrupolar correlation
and the specific heat. Neutron scattering measurement can be quite helpful to
probe both the low-energy modes like the Halperin-Saslow modes and
the spin correlation in different temperature regimes~\cite{PhysRevB.91.174402}. 
Nuclear magnetic resonance and
muon spin resonance experiments can also be useful to reveal the
dynamical properties of the system at different temperatures. On
the theoretical side, it would be interesting to establish a general
understanding and a phase diagram of a spin-2 model with both Heisenberg
and biquadratic interactions on the triangular lattice.

\emph{Acknowledgments.}---This work is financially supported by the National Natural Science Foundation of China under granting nos: No. 51532010, No. 91422303, No. 51472266 and No.2016YFA0301001 (GC); and the DOE, Basic Energy Sciences Grant No. DE-FG02-99ER45747 (ZQW).

\bibliography{refs}

\end{document}